\documentstyle[prd,epsf,aps,floats]{revtex}

\newcommand{\nn}{\nonumber}

\newcommand{\eqcm}{\: ,}   
\newcommand{\eqpt}{\: .}   

\newcommand{\ie}{{\it i.e.}}
\newcommand{\eg}{{\it e.g.}}
\newcommand{\cf}{{\it cf.\ }}
\newcommand{\etal}{{\it et al.}}

\newcommand{\eq}[1]{Eq.~(\ref{#1})}

\newcommand{\as}{\alpha_s}
\newcommand{\lqcd}{\Lambda_{QCD}}

\newcommand{\dpt}{\Delta_\perp}

\newcommand{\PL}[3]{Phys.\ Lett.\ {{\bf#1}} ({#3}) {#2}}
\newcommand{\NP}[3]{Nucl.\ Phys.\ {{\bf#1}} ({#3}) {#2}}
\newcommand{\PR}[3]{Phys.\ Rev.\  {{\bf#1}} ({#3}) {#2}}
\newcommand{\PRL}[3]{Phys.\ Rev.\ Lett.\ {{\bf#1}} ({#3}) {#2}}
\newcommand{\ZP}[3]{Z. Phys.\ {{\bf#1}} ({#3}) {#2}}
\newcommand{\PRT}[3]{Phys.\ Rept.\ {{\bf#1}} ({#3}) {#2}}
\newcommand{\SHEP}[3]{Surveys High Energy Phys.\ {{\bf#1}} ({#3})
  {#2}}

\begin{document}

\twocolumn[\hsize\textwidth\columnwidth\hsize\csname @twocolumnfalse\endcsname

\title{
\hbox to\hsize{\normalsize\hfil\rm DESY 98-189}
\hbox to\hsize{\normalsize\hfil\rm LAPTH 708/98}
\hbox to\hsize{\normalsize\hfil\rm NORDITA-98/57 HE}
\hbox to\hsize{\normalsize\hfil\rm SLAC-PUB-8015}
\hbox to\hsize{\normalsize\hfil hep-ph/9812277}
\vskip 40pt
Semi-Exclusive Processes: New Probes of Hadron Structure}

\author{Stanley J. Brodsky}
\address{Stanford Linear Accelerator Center, Stanford CA 94309, USA}

\author{Markus Diehl}
\address{Deutsches Elektronen-Synchroton DESY, D-22603 Hamburg,
  Germany}

\author{Paul Hoyer}
\address{Nordita, Blegdamsvej 17, DK-2100 Copenhagen, Denmark}

\author{St\'ephane Peign\'e}
\address{LAPTH/LAPP, F-74941 Annecy-le-Vieux Cedex, France}

\maketitle

\vskip2.0pc
\begin{abstract}
  We define and study hard ``semi-exclusive'' processes of the form
  $A+B \to C + Y$ which are characterized by a large momentum transfer
  between the particles $A$ and $C$ and a large rapidity gap between
  the final state particle $C$ and the inclusive system $Y$.  Such
  reactions are in effect generalizations of deep inelastic lepton
  scattering, providing novel currents which probe specific quark
  distributions of the target $B$ at fixed momentum fraction. We give
  explicit expressions for photo- and leptoproduction cross sections
  such as $\gamma p \to \pi Y$ in terms of parton distributions in the
  proton and the pion distribution amplitude. Semi-exclusive processes
  provide opportunities to study fundamental issues in QCD, including
  odderon exchange and color transparency, and suggest new ways to
  measure spin-dependent parton distributions.
\end{abstract}
\vskip2.0pc]

In this letter we shall study a new class of hard ``semi-exclusive''
processes of the form $A+B \to C + Y$, characterized by a large
momentum transfer $t= (p_A-p_C)^2$ and a large rapidity gap between
the final state particle $C$ and the inclusive system $Y$. Here $A, B$
and $C$ can be hadrons or (real or virtual) photons.  The cross
sections for such processes factorize in terms of the distribution
amplitudes of $A$ and $C$ and the parton distributions in the target
$B$. Because of this factorization semi-exclusive reactions provide a
novel array of generalized currents, which not only give insight into
the dynamics of hard scattering QCD processes, but also allow
experimental access to new combinations of the universal quark and
gluon distributions.

The hard QCD processes which have been mostly studied to date can be
divided into two main categories:
\begin{enumerate}
\item {\em Inclusive processes} such as DIS, $e p \to e + X$. In the
  limit of large photon virtuality $Q^2$ and energy $\nu$ in the
  target rest frame, the cross section can be expressed in terms of
  universal quark and gluon distributions $q(x,Q^2)$, $g(x,Q^2)$ in
  the target, where $x$ is the fraction of target momentum carried by
  the struck parton and $Q^2$ the factorization scale.
\item {\em Exclusive processes} such as $e p \to e p$. For large $Q^2$
  the form factor of the proton can be expressed in terms of its
  distribution amplitude, given by the valence Fock state wave
  function in the limit of vanishing transverse separation between the
  quarks~\cite{BL}.
\end{enumerate}

More recently it has been shown that the QCD scattering amplitude for
deeply virtual exclusive processes like Compton scattering $\gamma^* p
\to \gamma p$ and meson production $\gamma^* p \to M p$ factorizes
into a hard subprocess and soft universal hadronic matrix
elements~\cite{JiRad,CFS}. For example consider exclusive meson
electroproduction such as $e p \to e \pi^+ n$ (Fig.~1a). Here one
takes (as in DIS) the Bjorken limit of large photon virtuality , with
$x_B = Q^2/(2 m_p \nu)$ fixed, while the momentum transfer $t =
(p_p-p_n)^2$ remains small. These processes involve `skewed' parton
distributions, which are generalizations of the usual parton
distributions measured in DIS.  The skewed distribution in Fig.~1a
describes the emission of a $u$-quark from the proton target together
with the formation of the final neutron from the $d$-quark and the
proton remnants. As the subenergy $\hat s$ of the scattering process
$\gamma^* u \to \pi^+ d$ is not fixed, the amplitude involves an
integral over the $u$-quark momentum fraction $x$.
\begin{figure*}
\begin{center}
  \leavevmode
  \epsfxsize=3.5in
 \epsfbox{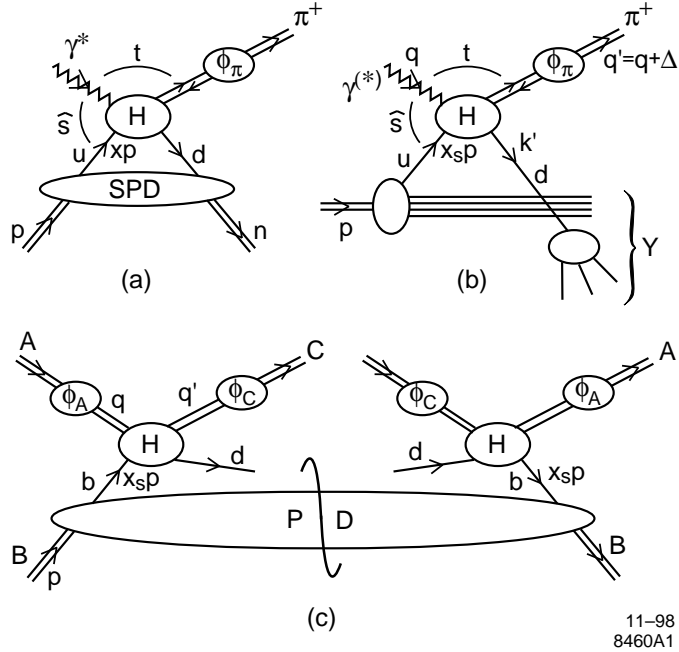}
\end{center}
\caption{{} (a): Factorization of $\gamma^* p \to \pi^+ n$ into a
  skewed parton distribution (SPD), a hard scattering $H$ and the pion
  distribution amplitude $\phi_\pi$. (b): Semi-exclusive process
  $\gamma^{(*)} p \to \pi^+ Y$. The $d$-quark produced in the hard
  scattering $H$ hadronizes independently of the spectator partons in
  the proton. (c): Diagram for the cross section of a generic
  semi-exclusive process. It involves a hard scattering $H$,
  distribution amplitudes $\phi_A$ and $\phi_C$ and a parton
  distribution (PD) in the target $B$.}
\end{figure*}

An essential condition for the factorization of the deeply virtual
meson production amplitude of Fig.\ 1a is the existence of a large
rapidity gap between the produced meson and the neutron. In fact,
this factorization remains valid if the neutron is replaced with a
hadronic system $Y$ of invariant mass $M_Y^2 \ll W^2$, where $W$ is
the c.m.\ energy of the $\gamma^* p$ process.

For $M_Y^2 \gg m_p^2$ the momentum $k'$ of the $d$-quark in Fig.~1b is
large with respect to the proton remnants, and hence it forms a jet.
This jet hadronizes independently of the other particles in the final
state if it is not in the direction of the meson, \ie, if the meson
has a large transverse momentum $q'_\perp = \Delta^{\phantom .}_\perp$
with respect to the photon direction in the $\gamma^* p$ c.m. Then the
cross section for an inclusive system $Y$ can be calculated as in DIS,
by treating the $d$-quark as a final state particle.

The large $\dpt$ furthermore allows only transversally compact
configurations of the projectile $A$ to couple to the hard subprocess
$H$ of Fig.~1b, as it does in exclusive processes~\cite{BL}. Hence the
above discussion applies not only to incoming virtual photons at large
$Q^2$, but also to real photons $(Q^2=0)$ and in fact to any hadron
projectile.

Let us then consider the general process $A+B\to C+Y$, where $B$ and
$C$ are hadrons or real photons, while the projectile $A$ can also be
a virtual photon. In the semi-exclusive kinematic limit
\begin{equation}
\lqcd^2,\, M_B^2,\, M_C^2 \ll
M_Y^2,\, \dpt^2 \ll W^2 \label{ylim}
\end{equation}
we have a large rapidity gap
\begin{equation}  \label{gap}
|y_C - y_d| = \log \frac{W^2}{\dpt^2 + M_Y^2}
\end{equation}
between $C$ and the parton $d$ produced in the hard scattering (see
Fig.~1c). The cross section then factorizes into the form
\begin{eqnarray}
\lefteqn{ \frac{d\sigma}{dt\,dx_S}(A+B\to C+Y) }
  \hspace{4em} \nonumber \\
&=& \sum_{b} f_{b/B}(x_S,\mu^2) \frac{d\sigma}{dt} (A b \to C d)
    \eqcm \label{gencross}
\end{eqnarray}
where $t=(q-q')^2$ and $f_{b/B}(x_S,\mu^2)$ denotes the distribution
of quarks, antiquarks and gluons $b$ in the target $B$. The momentum
fraction $x_S$ of the struck parton $b$ is fixed by kinematics to the
value
\begin{equation}
x_S = \frac{-t}{M_Y^2-t} \eqcm \label{xs}
\end{equation}
and the factorization scale $\mu^2$ is characteristic of the hard
subprocess $A b \to C d$.

In the kinematic limit (\ref{ylim}) the subenergy $\hat s = (q+x_S
p)^2$ of the hard process, the momentum transfer $t$, and the fraction
$x_F$ of the light-cone momentum of projectile $A$ carried by particle
$C$ are respectively given by
\begin{eqnarray}
\hat s &=& \frac{\dpt^2}{\dpt^2 + M_Y^2}\, W^2 \eqcm \label{shat} \\
-t     &=& \frac{\dpt^2+x_B M_Y^2}{1-x_B} \,=\,
           \frac{\dpt^2}{1-x_B/x_S} \eqcm \label{texpr} \\
1-x_F  &=& \frac{\dpt^2 + M_Y^2}{W^2} \eqcm \label{xf}
\end{eqnarray}
where we notice that $x_F$ is close to 1. We also have the relation
\begin{equation}
x_B \,<\, x_S = \frac{\dpt^2+x_B M_Y^2}{\dpt^2+ M_Y^2}
    \,<\, 1  \eqcm \label{xsbis}
\end{equation}
with $x_B = 0$ in the case where the projectile $A$ is a hadron or
real photon.

It is conceptually helpful to regard the hard scattering amplitude $H$
in Fig.~1c as a generalized current of momentum $q-q'=p_A - p_C$,
which interacts with the target parton $b$. For $A= \gamma^*$ we
obtain a close analogy to standard DIS when particle $C$ is removed.
With $q' \to 0$ we thus find $-t \to Q^2$, $M_Y^2 \to W^2$, and see
that $x_S$ in (\ref{xs}) goes over into $x_B = Q^2 /(W^2 + Q^2)$. The
possibility to control the value of $q'$ (and hence the momentum
fraction $x_S$ of the struck parton) as well as the quantum numbers of
particles $A$ and $C$ should make semi-exclusive processes a versatile
tool for studying hadron structure. The cross section further depends
on the distribution amplitudes $\phi_A$, $\phi_C$ (\cf Fig.~1c),
allowing new ways of measuring these quantities. The use of this new
current requires a sufficiently high c.m.\ energy, since according to
\eq{ylim} we need to have at least one intermediate large scale.  We
note that the possibility of creating effective currents using
processes similar to the ones we discuss here was considered already
before the advent of QCD \cite{BB}.

It is instructive to compare our semi-exclusive limit (\ref{ylim}) for
electroproduction, $Q^2 \sim W^2$, with the $x_F \to 1$ limit of
semi-inclusive DIS. After being struck by the virtual photon the
$u$-quark in Fig.~1b has a virtuality $\hat s \sim Q^2$ when
$\Delta_\perp^2 \sim M_Y^2$, cf.\ \eq{shat}. Since the time scale
$1/Q$ of the photon interaction is then similar to the time scale
$1/\sqrt{\hat s}$ of the further interactions of the struck quark,
these processes cannot be physically separated. Hence the hard
subprocess $H$ of Fig.~1b is compact. On the other hand, if
$\Delta_\perp^2 \ll M_Y^2$ the virtual photon time scale is much
shorter than that of quark fragmentation, and $H$ factorizes into
$\gamma^*u \to u$ times $u \to \pi^+ d$. This is the physics of
semi-inclusive DIS and also of lepton pair production ($\pi p \to
\mu^+ \!\mu^- Y$) in the limit $x_F \to 1$ when $\Delta_\perp$ is
integrated over~\cite{DY}.

Pion photoproduction at large transverse momentum $\Delta_\perp$ was
studied in Ref.\ \cite{acw} for $x_F < 1$, \ie, in the case of no
rapidity gap. In this case the struck quark emits a gluon at a short
time-scale $1/\Delta_\perp$, but the pion is predominantly produced
via a standard non-perturbative fragmentation process.

Next we consider in more detail the specific semi-exclusive process
$\gamma^{(*)} p \to \pi^+ Y$ shown in Fig.~1b. We work in the
kinematic limit (\ref{ylim}) and for simplicity take a single
intermediate scale, $\dpt^2 \sim M_Y^2$. The virtuality $Q^2$ of the
photon can scale as
\begin{equation}
Q^2 \sim \left\{
\begin{array}{cll}
0     & \hspace{1em} & \mbox{photoproduction} \\
M_Y^2 & \hspace{1em} & \mbox{DIS, } x_B \to 0 \\
W^2   & \hspace{1em} & \mbox{DIS, } x_B \mbox{ finite}
\end{array} \right. \label{qlim}
\end{equation}
Note that according to Eqs.~(\ref{shat}) and (\ref{texpr}) $-t \sim
\dpt^2$ is of intermediate scale, and $\hat{s} \sim W^2$ is very
large, so that we have $\lqcd^2 \ll -t \ll \hat{s}$ in the hard
scattering.

The target parton ($b$ in Fig.~1c) attached to the hard amplitude $H$
can (at lowest order in $\as$) be either one of the valence quarks
$u,\bar d$ of the $\pi^+$. These two contributions add incoherently in
the cross section, weighted by the respective parton distribution
$f_{b/p}$. In Fig.~2a we show one the four diagrams which contribute to
$H$ in the case $b=u$. The three other diagrams are obtained by
different orderings of the photon and gluon vertices on the $u$- and
$d$-quark lines.

\begin{figure}
\begin{center}
  \leavevmode
  \epsfxsize 0.45\textwidth
  \epsfbox{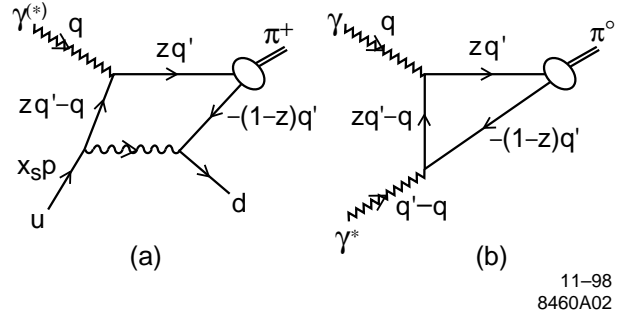}
\end{center}
\caption{{} (a): One of the diagrams of the hard scattering
  $\gamma^{(*)} u \to \pi^+ d$ to leading order in $\as$. (b): One of
  the leading order diagrams for the pion transition form factor in
  $\gamma^* \gamma \to \pi$.}
\end{figure}

Exclusive processes can be sensitive to infrared end-point
contributions, where the momentum of one of the valence quarks in a
hadron wave function vanishes. In Fig.~2a it may be seen that the
gluon propagator goes on-shell for $z \to 1$, since its momentum then
equals that of the final $d$-quark.  It is important to note that the
gluon four-momentum does not vanish in this limit, since it still has
a large transverse component $-\vec\dpt$. As a consequence the
$u$-quark propagator does not become singular at this end-point, in
other words $(zq'-q)^2 = zt-(1-z)Q^2$ is ``protected'' by the large
momentum transfer $-t$ in the $z \to 1$ limit. This suggests that the
hard amplitude is no more sensitive to end-point contributions than
the pion transition form factor (Fig.~2b). On the other hand, in
exclusive meson production at large $Q^2$ and small $t$ both internal
propagators in Fig.~2a go on-shell at $z=1$, which is what makes the
amplitude with transversely polarized virtual photons infrared
sensitive \cite{CFS}.

We also note that (for $z \neq 0, 1$)
all propagators in the hard scattering subprocess have at least a
virtuality of order $Q^2$ or $-t$, whichever is larger.
This ensures that the scattering amplitude $H$ is compact, and that
the photon couples coherently to both valence quarks (\ie, all four
diagrams contribute at leading order). In contrast to ordinary DIS and
semi-inclusive processes, the contribution of the parton distribution
$f_{u/p}$ will thus not necessarily be weighted by $e_u^2$, the square
of the electric charge of the struck quark.

In the case of photoproduction one finds in the limit of \eq{ylim}
with $Q^2=0$ that the $\gamma u \to \pi^+ d$ subprocess cross section
is \cite{acw}
\begin{eqnarray}
\lefteqn{\frac{d\sigma}{dt}(\gamma u \to \pi^+ d) =
          \frac{128\pi^2}{27}\, \alpha\as^2\,
          \frac{(e_u-e_d)^2}{{\hat s}^2 (-t)} } \hspace{3em} \nn \\
&\times& \left\{\left[\int_0^1 dz \frac{\phi_\pi(z)}{z} \right]^2 +
\left[\int_0^1 dz \frac{\phi_\pi(z)}{1-z} \right]^2 \right\} \eqcm
  \label{photocross}
\end{eqnarray}
where we choose the convention of Refs.\ \cite{BL,acw} for the pion
distribution amplitude, namely $\int dz\, \phi_\pi(z) = f_\pi /
\sqrt{12}$ with $f_\pi = 93$ MeV. The $\gamma \bar d \to \pi^+ \bar u$
subprocess has an identical cross section. The result for the physical
process $\gamma p \to \pi^+ Y$ is then, according to \eq{gencross},
\begin{eqnarray}
\lefteqn{ \frac{d\sigma}{dt\,dx_S}(\gamma p \to \pi^+ Y) }
  \hspace{2em} \nn \\
&=& \left[u(x_S,-t) + \bar d(x_S,-t) \right]
    \frac{d\sigma}{dt}(\gamma u \to \pi^+ d) \label{fullphoto}
\end{eqnarray}
with the notation $q(x_S,-t) = f_{q/p}(x_S,-t)$. There are several
interesting aspects of this result.
\begin{itemize}
\item Both the target $u$- and $\bar d$-quark contributions are
  weighted by the {\em total charge} $e_u-e_d = +1$ of the produced
  $\pi^+$. An analogous formula holds of course if the $\pi^+$ is
  replaced with another pseudoscalar. For neutral meson production,
  $\gamma p \to M^0 Y$ with $M^0=\pi^0$, $K^0$, $\eta$, \ldots, the
  expression (\ref{photocross}) \emph{vanishes}; more exactly one
  finds that the cross section is suppressed by $(-t /\hat{s})^2$
  compared with the charged meson case~\cite{acw}.  This illustrates
  that the new type of current probe we are considering weights the
  target parton distributions differently from the photon current, as
  is also the case for weak currents.
\item The cross section \eq{photocross} has a power-law behavior, $d
  \sigma/ d t \propto 1 /\hat{s}^3$ at fixed $t /\hat{s}$. This is the
  basic signature that the amplitude factorizes into a meson
  distribution amplitude and a hard scattering subprocess.
\item At fixed $t$ the expression (\ref{photocross}) goes like $1
  /\hat s^2$, which is characteristic of two spin 1/2 quark exchanges
  in the $t$-channel. Notice that with $\lqcd^2 \ll -t \ll \hat{s}$
  the hard scattering takes place in the perturbative Regge regime: if
  a deviation from this $\hat{s}$-behavior were to be observed
  experimentally it would indicate that the quark exchange reggeizes,
  \ie, that contributions from higher order ladder diagrams are
  important. For a discussion of QCD expectations and experimental
  evidence on how meson Regge trajectories $\alpha(t)$ behave at large
  $-t$ we refer to~\cite{BTT}. Let us emphasize that perturbative
  Reggeization would not change the power-law behavior in $\hat{s}$ at
  fixed $t /\hat{s}$ mentioned above.
\item The pion distribution amplitude $\phi_\pi(z)$ enters in
  precisely the same way as it does in the pion transition form factor
  for $\gamma^* \gamma \to \pi^0$ \cite{BL,acw}
\begin{equation}
F_{\pi\gamma}(Q^2) = {\sqrt{48}\,
(e_u^2- e_d^2) \over Q^2} \int^1_0 dz {\phi_\pi(z)\over z}  \eqpt
\label{fpigam}
\end{equation}
A comparison of Figs.~2a and 2b suggests that this may again be
interpreted as the result of replacing the $\gamma^*$ probe with an
effective $u\bar d$ current of hardness $t$. Hence the relation
between observables,
\begin{equation}
\frac{d\sigma}{dt}(\gamma u \to \pi^+ d) = {16 \pi^2 \over 9}\, \alpha
\alpha_s^2\, {-t \vert F_{\pi\gamma}(-t)\vert^2\over {\hat s}^2} \eqcm
\label{fpigamrel}
\end{equation}
which holds at lowest order in $\as$,
may have a broader range of validity. Note that in order to obtain
\eq{fpigamrel} we have used isospin symmetry, $\phi_{\pi^+}(z) =
\phi_{\pi^0}(z)$, which also implies $\phi_\pi(z)= \phi_\pi(1-z)$.
\end{itemize}

In the limit (\ref{ylim}) with $Q^2 \sim W^2$, \ie, at finite $x_B$,
the semi-exclusive electroproduction cross section is
\begin{eqnarray}
\lefteqn{ \frac{d\sigma(ep\to e\pi^+Y)}{dQ^2\,dx_B\,dt\,dx_S} =
\frac{\alpha}{\pi} \frac{1-y}{Q^2 x_B}\,
\frac{512\pi^2}{27}\, \alpha\as^2
\frac{x_B}{\hat{s}\, Q^4 x_S}} \hspace{0.5em} \nonumber\\
&&
\times \left[\int_0^1 dz \frac{\phi_\pi(z)}{z} \right]^2
\left\{ u(x_S) \left[e_u+
\Big(1-\frac{x_B}{x_S}\Big)\, e_d \right]^2 \right.
\nonumber \\
&& \hspace{7.7em} + \left.
\bar d(x_S) \left[e_d+
\Big(1-\frac{x_B}{x_S}\Big)\, e_u \right]^2 \right\}
\label{elcross}
\end{eqnarray}
where $y=\nu/E_e$ is the momentum fraction of the projectile electron
carried by the virtual photon, and we have used again $\phi_\pi(z)=
\phi_\pi(1-z)$. We make the following remarks.
\begin{itemize}
\item The semi-exclusive cross section in \eq{elcross} corresponds to
  longitudinal photon exchange. The contribution from transverse
  photons is suppressed, as in the exclusive case $\gamma^* p \to Mp$
  at large $Q^2$ and small $-t$~\cite{CFS}.
\item For $\Delta_\perp^2 \ll M_Y^2$ we have $x_S \to x_B$ according
  to \eq{xsbis}. We find that the parton distributions are then
  multiplied by the corresponding quark charge squared in the cross
  section. This is a consequence of the fact that, as discussed above,
  the hard subprocess factorizes in this limit into a virtual photon
  interaction and a quark fragmentation process. We note that
  \eq{elcross} was derived for $\dpt^2 \sim M_Y^2$ and acquires
  corrections when $\Delta_\perp^2 \ll M_Y^2$.
\item The subprocess is of hardness $Q^2$, which thus is the relevant
  scale for the quark parton distributions. If one takes into account
  higher orders in $\as$ the target parton couples to a ladder. Its
  hardness, which then becomes the appropriate scale, will be between
  $-t$ and $Q^2$.
\end{itemize}

In the intermediate range $Q^2 \sim M_Y^2$ of \eq{qlim} transverse and
longitudinal photon polarizations contribute with comparable strength,
and the structure of the cross section is richer than in the two
extreme cases just discussed.

We conclude with a number of more general remarks and suggestions for
future work.

{\em 1. Vector mesons.} In addition to pseudoscalar mesons one can
also consider vector meson production. We find that exact analogs of
Eqs.~(\ref{photocross}) and (\ref{elcross}) hold if the vector meson
is longitudinally polarized. Transverse vector mesons are suppressed
in the cross section by $(- t /\hat{s})^2$ for photo- and $-t /
\hat{s}$ for electroproduction. For symmetry reasons there is no
interference between different meson polarizations.

{\em 2. Particle production ratios.} Systematic comparisons of
semi-exclusive photoproduction of various particles can give useful
information on parton distributions and distribution amplitudes. The
hard subprocess (\ref{photocross}) cancels in the ratio of physical
cross sections (\ref{fullphoto}) for $\pi^+$ and $\pi^-$. Hence
$d\sigma(\pi^+)/d\sigma(\pi^-)$ directly measures the $(u+\bar
d)/(d+\bar u)$ parton distribution ratio. Similarly, the
$d\sigma(K^+)/d\sigma(K^-)$ ratio measures the strange quark content
of the target without uncertainties due, \eg, to fragmentation
functions.

Conversely, the parton distributions drop out in the ratio
$d\sigma(\rho_L^+)/d\sigma(\pi^+)$ of longitudinally polarized $\rho$
mesons to pions, allowing a comparison of their distribution
amplitudes. Since the normalization of both $\phi_\rho$ and $\phi_\pi$
is fixed by the leptonic decay widths such a comparison can reveal
differences in their $z$-dependence. In the intermediate $Q^2$-range
of \eq{qlim} the relative size of $Q^2$ and $t$ can furthermore be
tuned to change the dependence of the hard subprocess on $z$. This may
be used to get further information on the shape of $\phi(z)$.

{\em 3. Hard odderon and pomeron exchange.}  As we noted above, the
lowest order quark exchange contribution to $\gamma p \to \pi^0 Y$ is
strongly suppressed. At higher orders in $\as$ there is a contribution
from scattering on gluons ($b=g$ in Fig.\ 1c). Due to charge
conjugation at least two gluons need to be emitted from the hard
scattering ($d=gg$ in Fig.\ 1c). Altogether three gluons are thus
exchanged in the $t$-channel, corresponding to hard odderon exchange.
Semi-exclusive production of $\pi^0$, $\eta$ or other neutral
pseudoscalars may thus be particularly sensitive to odderon effects
since competing production mechanisms are suppressed. See
Ref.~\cite{LN} for a recent discussion of odderon physics.

An analogous argument suggests that $\gamma p \to \rho^0 Y$ is a good
process for studying hard two-gluon ladders, \ie, pomeron exchange at
large $t$. In this context neutral vector meson production has in fact
been studied in a kinematic region very similar to
ours~\cite{VMTh,VMExp}. Real photon production, $\gamma p \to \gamma
Y$, may also be interesting in this respect~\cite{IvWu}, since
compared to two-gluon exchange the quark exchange contribution is
again suppressed by a power of $-t /\hat s$.

{\em 4. Spin and transversity distributions.} Polarization of the
target $B$ can naturally be incorporated in our framework. A
longitudinally polarized target selects the usual spin-dependent
parton distributions $\Delta q(x_S)$. It also appears possible to
measure the quark transverse spin, or transversity distribution in
photoproduction of $\rho$ mesons on transversely polarized protons. In
this case only the interference term between transversely and
longitudinally polarized $\rho$ mesons should contribute. Since very
little experimental information on the transversity distribution is
available, this question merits further study.

{\em 5. Color transparency.} The factorization of the hard amplitude
$H$ in Fig.\ 1c from the target remnants is a consequence of the high
transverse momentum which selects compact sizes in the projectile $A$
and in the produced particle $C$. In the case of nuclear targets this
color transparency \cite{CT} implies according to \eq{gencross} that
all target dependence enters via the nuclear parton distribution. Thus
tests of color transparency can be made even in photoproduction, \eg,
through
\begin{equation}
\gamma A \to \left\{
\begin{array}{c}
\pi^+(\dpt) + Y \\
p(\dpt) + Y
\end{array} \right. \label{ctest}
\end{equation}
in the semi-exclusive kinematic region specified by \eq{ylim}.

Color transparency has so far been studied mainly in exclusive
processes where the target scatters elastically, such as $\gamma^*A
\to \rho A$ \cite{E665}, $pA \to pp+(A-1)$ \cite{BNL} and $\gamma^*A
\to p+(A-1)$ \cite{NE18}. Semi-exclusive processes provide a
possibility to study color transparency in processes where the target
dissociates into a heavy inclusive system $Y$ \cite{PH}. This puts
less stringent requirements on the energy resolution of the apparatus,
but it also requires a higher beam energy to ensure the existence of a
rapidity gap.

\section*{Acknowledgments}
It is a pleasure to acknowledge discussions with B. Pire and O.
Teryaev.

We are grateful for the hospitality of the European Centre for
Theoretical Studies in Nuclear Physics and Related Areas (ECT*), where
part of this work was done. MD and PH also wish to thank for the
hospitality of CPhT, Ecole Polytechnique. SJB is supported in part by
the Department of Energy, contract DE-AC03-76SF00515, and MD, PH and
SP are supported in part by the EU/TMR contract EBR FMRX-CT96-0008.


\begin{references}
  
\bibitem{BL} S. J. Brodsky and G. P. Lepage, \PR{D22}{2157}{1980} and
  \PR{D24}{1808}{1981}. For reviews and further references, see S. J.
  Brodsky and G. P. \mbox{Lepage}, in: Perturbative Quantum
  Chromodynamics, ed.\ A.~H. Mueller (World Scientific, Singapore,
  1989); V. L.  Chernyak and A. R. Zhitnitsky, \PRT{122}{173}{1984}.


\bibitem{JiRad}
  X. Ji, \PR{D55}{7114}{1997}, hep-ph/9609381; \\
  X. Ji and J. Osborne, \PR{D58}{094018}{1998}, hep-ph/9801260; \\
  A.V. Radyushkin, \PR{D56}{5524}{1997}, hep-ph/9704207.

\bibitem{CFS}
J. C. Collins, L. Frankfurt and M. Strikman,
  \PR{D56}{2982}{1997}, hep-ph/9611433.

\bibitem{BB}
  J. F. Gunion, S. J. Brodsky and R. Blankenbecler, \PR{D6}{2652}{1972}; \\
  R. Blankenbecler and S. J. Brodsky, \PR{D10}{2973}{1974}.

\bibitem{DY}
  E. L. Berger and S. J. Brodsky, \PRL{42}{940}{1979}; \\
  A. Brandenburg, V. V. Khoze and D. M\"uller, \PL{B347}{413}{1995},
  hep-ph/9410327.

\bibitem{acw}
  C. E. Carlson and A. B. Wakely, \PR{D48}{2000}{1993}; \\
  A. Afanasev, C. E. Carlson and C. Wahlquist, \PL{B398}{393}{1997},
  hep-ph/9701215 and \PR{D58}{054007}{1998}, hep-ph/9706522.

\bibitem{BTT}
S. J. Brodsky, W.-K. Tang and C. B. Thorn, \PL{B318}{203}{1993}.

\bibitem{LN}
P. V. Landshoff and O. Nachtmann, hep-ph/9808233.

\bibitem{VMTh}
  L. Frankfurt and M. Strikman, \PRL{63}{1914}{1989}; \\
  H. Abramowicz, L. Frankfurt and M. Strikman, \SHEP{11}{51}{1997},
  hep-ph/9503437; \\
  J. R. Forshaw and M. G. Ryskin, \ZP{C68}{137}{1995}, hep-ph/9501376; \\
  J. Bartels, J. R. Forshaw, H. Lotter and M. W\"usthoff,
  \PL{B375}{301}{1996}, hep-ph/9601201.

\bibitem{VMExp}
H1 Collaboration, ``Production of $J/\Psi$ mesons with
  large $|t|$ at HERA'', contribution to the International Europhysics
  Conference on High Energy Physics (EPS97), Jerusalem,
  Israel, August 1997;
  ZEUS Collaboration, ``Study of vector meson production at large
  $|t|$ at HERA'', contribution to the International Europhysics
  Conference on High Energy Physics (EPS97), Jerusalem, Israel, August
  1997.

\bibitem{IvWu} D. Yu. Ivanov and M. W\"usthoff, hep-ph/9808455.

\bibitem{CT}
  S. J. Brodsky and A. H. Mueller, \PL{B206}{685}{1988}; \\
  L. Frankfurt and M. Strikman, \PRT{160}{235}{1988}; \\
  P. Jain, B. Pire and J. P. Ralston, \PRT{271}{67}{1996},
  hep-ph/9511333.

\bibitem{E665}
E665 Collaboration, M. R. Adams \etal,
  \PRL{74}{1525}{1995}.

\bibitem{BNL}
R. S. Carroll \etal, \PRL{61}{1698}{1988}.

\bibitem{NE18}
NE-18 Collaboration, N. C. R. Makins \etal,
  \PRL{72}{1986}{1994}; \\
  T. G. O'Neill \etal, \PL{B351}{87}{1995}, hep-ph/9408260.

\bibitem{PH}
P. Hoyer, \NP{A622}{284c}{1997}, hep-ph/9703462.

\end{references}
\end{document}